# BxC Toolkit: Generating Tailored Turbulent 3D Magnetic Fields

Daniela Maci[1], Rony Keppens[1], and Fabio Bacchini[1,2]
[1] Centre for mathematical Plasma Astrophysics, Department of Mathematics, KU Leuven, Celestijnenlaan 200B, B-3001 Leuven, Belgium; daniela.maci@kuleuven.be
[2] Royal Belgian Institute for Space Aeronomy, Solar-Terrestrial Centre of Excellence, Ringlaan 3, 1180 Uccle, Belgium

## Abstract

Turbulent states are ubiquitous in plasmas, and the understanding of turbulence is fundamental in modern astrophysics. Numerical simulations, which are the state-of-the-art approach to the study of turbulence, require substantial computing resources. Recently, attention shifted to methods for generating synthetic turbulent magnetic fields, affordably creating fields with parameter-controlled characteristic features of turbulence. In this context, the BxC toolkit was developed and validated against direct numerical simulations (DNSs) of isotropic turbulent magnetic fields. Here, we demonstrate novel extensions of BxC to generate realistic turbulent magnetic fields in a fast, controlled, geometric approach. First, we perform a parameter study to determine quantitative relations between the BxC input parameters and the desired characteristic features of the turbulent power spectrum, such as the extent of the inertial range, its spectral slope, and the injection and dissipation scale. Second, we introduce in the model a set of structured background magnetic fields, $\boldsymbol{B}_0$, as a natural and more realistic extension to the purely isotropic turbulent fields. Third, we extend the model to include anisotropic turbulence properties in the generated fields. With all these extensions combined, our tool can quickly generate any desired structured magnetic field with controlled, anisotropic turbulent fluctuations, faster by orders of magnitude with respect to DNSs. These can be used, e.g., to provide initial conditions for DNSs or easily generate synthetic data for many astrophysical settings, all at otherwise unaffordable resolutions.

*Unified Astronomy Thesaurus concepts:* Space plasmas (1544); Plasma astrophysics (1261); Magnetohydrodynamics (1964)

## 1. Introduction

Fluids, gases, and plasmas are often found in turbulent states of motion, making the study of turbulence a central topic in many research fields, from hydrodynamics to modern astrophysics (Biskamp 2003; Brandenburg & Lazarian 2013; Galtier 2016; Goedbloed et al. 2019). The importance of turbulence largely comes from the ubiquitous presence of such a state of motion. The present work focuses on (but is not restricted to) astrophysical applications, where turbulence is observed in numerous environments: in accretion disks and astrophysical jets, in solar/stellar atmospheres and winds, in molecular clouds or galaxies, etc. (Barnes 1979; Parker 1979; Balbus & Hawley 1998; Schekochihin & Cowley 2007; Beresnyak & Lazarian 2019). Nonetheless, this work builds on an issue that is shared by all turbulence-related fields of study. The state-of-the-art approach to study turbulent states is to perform direct numerical simulations (DNSs), which are effective yet often extremely expensive in terms of the computational resources required.

With the aim of reducing the need for expensive numerical simulations, recent studies (Juneja et al. 1994; Cametti et al. 1998; Zimbardo et al. 2000; Ruffolo et al. 2006; Malara et al. 2016; Lübke et al. 2023, 2024) have been focusing on the development of software that can generate synthetic data of turbulent quantities, e.g., magnetic fields. The general approach used in synthetic turbulence is to avoid solving physical equations numerically, instead using simplified models and algorithms that are able to mimic properties that are characteristic of turbulence. In this paper, we develop an alternative synthetic model based on the previously presented BxC code (Durrive et al. 2022), which is intended as a general, versatile model, not restricted to any specific application. BxC, which stands for magnetic field from multiplicative chaos, is a code fully implemented in Python that can rapidly and inexpensively produce data cubes of turbulent magnetic fields of order $\sim 1000^3$ points and more. The present work is based on a double intent: having an easily customizable turbulent power spectrum and giving users the possibility to generate fields that are closer to real astrophysical scenarios by introducing physical characteristics, such as anisotropy, while keeping the model as simple and efficient as possible.

In almost all synthetic models, great importance is given to the power spectrum of turbulent fields, which is typically expected to show a power-law decay in a range of values that extends from the injection scale to the dissipation scale (i.e., in the inertial range). This concept was formalized by Kolmogorov's theory (Kolmogorov 1991), and it is now established in the literature as a distinctive characteristic of turbulence. In addition to the shape of the power spectrum, distinctive properties such as anisotropy and intermittency should be input-controlled in any synthetic model. The concept of anisotropy is strictly related to the presence of a guide field, which causes differences in the parallel and perpendicular directions in terms of scaling laws and hence energy transfer mechanisms. Intermittency is a property related to the statistical distribution of spatial increments of the fields. In practice, in a turbulent field, one would expect the distribution to show heavily non-Gaussian tails for small increments in the magnetic field $\boldsymbol{B}$ while tending to Gaussianity for large increments. BxC convincingly mimics and reproduces the intermittent and non-Gaussian character of a turbulent magnetic field, as demonstrated in Durrive et al. (2022), where fields generated by BxC have been successfully compared and validated against results from a magnetohydrodynamic DNS. The two fields (the one generated by BxC and the one obtained with the DNS) show the same properties in terms of power spectrum, probability distribution function (PDF) of increments (in both the magnetic field $B$ and the current density $j$), structure functions, and spectrum of exponents, hence proving the turbulent and intermittent character of the fields generated by BxC. In addition to showing power-law decaying power spectra, intermittency, and anisotropy, which are properties that may be more or less important according to the specific application for which the model is designed, synthetic algorithms can differ from each other in terms of spatial dimensionality (e.g., 1D, 2D, or 3D) and whether the fields are time-dependent or not. Although we primarily focus on static fields, Durrive et al. (2022) showed how BxC also has the potential of reproducing a time-dependent magnetic field evolution by continuously varying the input control parameters.

Given the many features to take into consideration, a variety of models have been proposed so far. In this context, two classes of synthetic models have attracted widespread attention: wavelet-based algorithms (e.g., Juneja et al. 1994; Cametti et al. 1998; Malara et al. 2016), which are very efficient in reproducing intermittency in the fields, and Fourier-based sampling procedures (e.g., Zimbardo et al. 2000; Ruffolo et al. 2006), in which the model can reproduce anisotropy but intermittency is not present. Recently, a combination of both approaches has also been proposed by Lübke et al. (2023, 2024). BxC does not belong to either of these categories, but, as suggested by the name itself, it uses a completely alternative approach linked to the concept of *multiplicative chaos* (Kahane 2000a, 2000b; Durrive et al. 2020), which is a wider research field on (hydrodynamic) turbulence (see Rhodes & Vargas 2014 for a review of the topic). Indeed, BxC has been inspired by recent developments in hydrodynamics, which suggest a relation between random fields and incompressible hydrodynamic turbulent fields (Chevillard et al. 2011). At the same time, BxC differs from this approach in terms of the fundamental structures and visual aspects of the generated fields. Specifically, Chevillard et al. (2011) succeed in incorporating in the model mathematical properties of turbulence but lack a visual turbulent appearance in vortices and current sheets. BxC has been developed with the aim of also reproducing visual characteristics of turbulent magnetic fields, meaning that BxC-generated fields also have an actual "look-and-feel" resemblance to turbulent fields, such as hierarchically structured magnetic eddies and current sheets. This is because the actual nonlinear transformation of fractional Gaussian fields (FGFs) in our toolkit builds in spiral shapes and multilayered current sheets, which are typical for all DNS realizations (see Section 2 for more details on the algorithm). So far, this aspect has been underexplored, and it fundamentally distinguishes BxC from Chevillard et al. (2011), as well as most other synthetic models. Indeed, to the best of our knowledge, no other synthetic turbulence model allows for the fully customizable production of turbulent fields including both structures and higher-order statistics at the same time.

Assuming the turbulent and intermittent character of the fields generated using BxC, in this paper, we introduce several novel capabilities into BxC to construct more realistic turbulent magnetic fields. Section 2 briefly describes the important features of the BxC model, including recent modifications that have been been introduced compared to the original algorithm (Durrive et al. 2022). Sections 3 and 4 contain the results of this study. In Section 3, we present the results of a parameter study conducted on the control parameters, which are varied independently from one another in order to assess and quantify their effects on the power spectrum. In Section 4, we generalize the model to reproduce more realistic scenarios. In particular, the isotropic fully turbulent fields generated by BxC are extended to turbulent fields with a background structure (Section 4.1) and to fields featuring anisotropic turbulence (Section 4.2). Finally, the conclusions of this study are presented in Section 5.

## 2. Methodology

The BxC toolkit is a Python implemented code that generates data cubes of turbulent magnetic fields using a combination of analytical formulae and geometric constructions. The work presented here builds on the preexisting algorithm, implementing new features that render the generated fields even more realistic. In order to contextualize the new features that are now included in the model, we first briefly summarize the relevant aspects of the BxC algorithm, as presented in Durrive et al. (2022).

The basic idea behind the BxC model is to start from a white-noise vector field and transform it in a strategic way in order to obtain magnetic and current vector structures with statistical properties that are typical of a turbulent magnetic field. The white-noise vector field is generated from a normal distribution with zero mean ($\mu = 0$) and unit standard deviation ($\sigma = 1$), but both $\mu$ and $\sigma$ are input parameters that can be user defined. The same value of the mean and standard deviation is used for all three vector components. As a first step, the white-noise vector field is used to generate an FGF, $R$, which, in turn, is used to generate the final turbulent magnetic field $B$. In both these steps, the fields are transformed analytically using a modified version of the Biot–Savart law. Before it is used to generate $B$, $R$ is subject to a nonlinear, geometrically inspired transformation. In the original formulation, the Biot–Savart law relates the magnetic field $B$ to the current density field $j$ through the convolution:

$$B \propto \int_{\mathbb{R}^3} \frac{j \times r}{r^3} dV. \tag{1}$$

The modified versions implemented in the BxC model are, instead:

$$R \propto \int_{r \leqslant L_R} \frac{\tilde{s} \times r}{(r^2 + \eta_R^2)^{h_R}} dV, \tag{2}$$

$$B \propto \int_{r \leqslant L_B} \frac{c(R) \times r}{(r^2 + \eta_B^2)^{h_B}} dV. \tag{3}$$

A clear difference between the standard Biot–Savart law and the ones implemented in the model is the replacement of the current density $j$ in Equation (2) with a white-noise vector $\tilde{s}$, which generates the FGF, $R$. In Equation (3), the current density is replaced by a nonlinear transformation of $R$, which thereby introduces deviations from Gaussianity and generates the turbulent magnetic field, $B$. We may denote this nonlinear transformation as $c = SR$, where $S$ is a geometrically controlled transcendental operator defined in terms of gradient and norm of $R$. The specific choice of $S$ is inspired by two visual characteristic features of turbulent fields: first, the omnipresence of spiral-shaped current sheets swirling around through space, and second, the appearance of intense sheets surrounded by more diffuse ones. The final formula for $S$ is

$$S(\lambda_R) = T_i(\lambda_R) + \varepsilon T_d(\cos(k_d \lambda_R)), \quad (4)$$

where $T_i$ and $T_d$ are top-hat functions that effectively reproduce the intense and diffuse sheets, respectively, and

$$\lambda_R = |R| - c_0 - d\theta_R \quad (5)$$

$$\text{with} \quad \theta_R = \frac{1}{\pi} \text{atan2}\left(\frac{\partial_y |R|}{\partial_x |R|}\right). \quad (6)$$

Equation (5) is a 3D generalization of an Archimedean spiral, in which the norm of the FGF $|R|$ plays the role of radius and $\theta_R$ (in Equation (6)) is defined accordingly in terms of the gradient of $|R|$ (i.e., it is interpreted as an angle). Equation (4) defines a scalar field dependent on $|R|$ that multiplies $R$ component-wise, hence preserving the vectorial nature of $c$.

In practice, the BxC algorithm does not artificially build the magnetic field directly, but it generates the turbulent magnetic field starting from a carefully constructed "current density" field, $c$. This is the key aspect of the process that eventually yields structures in the fields. The actual current density vector can be computed afterward from the relation $j = \nabla \times B$. In addition to the replacement of the current density, other changes are applied in order to replace Equation (1) with Equations (2) and (3): the domain is limited to cubes of side $L_R$ and $L_B$, respectively; the power-law behavior is extended to the free exponents $h_R$ and $h_B$; and finally, regularization is performed through $\eta_R$ and $\eta_B$ to avoid singularities at the origin. This process naturally introduces into the code a set of input parameters that will be the focus of the parameter study in the next section. It is also worth mentioning that building the magnetic field using Equation (3) has the intrinsic advantage of ensuring a divergence-free field, without additional constraints on $c$ (see the Appendix).

The magnetic fields obtained using the model described above are fully isotropic by construction, since there is no preferential direction imposed at any point. However, in many realistic astrophysical environments, magnetic turbulence is often found to develop anisotropy along some preferential direction, i.e., with respect to a background or guide magnetic field. With the purpose of reproducing turbulence in such environments, we detail here how the model is modified such that it allows for either structured background magnetic fields and/or the choice of a preferential direction. One of the main concerns in achieving this generalization is to retain the same algorithmic structure in order to maintain similar computational times and not overcomplicate the model itself. The most immediate way to achieve this direction-based variation of the fields is to exploit the input parameters of the model. In principle, almost all the input parameters present in the model can be adapted to become direction-dependent. However, the limited set of direction-dependent parameters discussed below was chosen upon consideration of what would physically render the fields anisotropic. In practice, having parameters that are direction-dependent means that instead of considering the same value in every direction, we now give as input three different values corresponding to the $x$-, $y$-, and $z$-directions, respectively. In particular, we will show (see Section 4.2) how anisotropy can already be controlled by introducing direction-dependent values of

1. the white-noise standard deviation ($\sigma$) and
2. the integration region for $R$ ($L_R$).

To recover isotropic fields with this new generalized approach, it is sufficient to set each parameter value equal in all directions, retrieving the original BxC algorithm (Durrive et al. 2022).

## 3. Parameter Study

One of the main advantages of BxC lies in the possibility to customize the generated fields not only from a visual point of view but also from a statistical one. In previous work, Durrive et al. (2022) already suggested this possibility, showing that variations in the input parameters are reflected in variations in the statistical aspects of the fields. It was demonstrated how parameter changes influence multiple aspects of the generated fields in terms of spectral properties, but the original paper (Durrive et al. 2022) did not provide specific relations between the model's input parameters and statistical properties. Therefore, in this work, we perform an in-depth, quantitative study on the influence of the input parameters on the final turbulent fields.

We vary each parameter independently from the others in order to assess its impact on the power spectrum. Each parameter has been limited to values that produce reasonable turbulent fields both visually and statistically. Figure 1 shows a typical 3D turbulent field generated by BxC. The figures in the left panels are all related to the magnetic field $B$, while the ones on the right show the current density field $j$. The top panels show the norm of the fields, which display a fluidlike behavior typical of turbulence. In the magnetic field picture, one can recognize vortex-like structures in which the intensity smoothly alternates from high to low, and vice versa. Similarly, the current density plot exhibits few high-intensity thin sheets in the shape of spirals, surrounded by more diffuse ones. The same visual aspects are confirmed in the middle row, in which we show 2D cuts (one for each Cartesian plane) of isocontours corresponding to different intensity levels. In the left panel of the bottom row, we show magnetic field lines, which confirm that the field is isotropic and there is no preferential direction. The right panel of the bottom row, instead, shows 3D isocontours of the current density field, confirming that the current density is indeed distributed in sheetlike structures and that high-intensity structures are very sparse and not volume filling.

Details on the ranges considered for each parameter can be found in Table 1, together with the fixed values used for each parameter when varying the others (i.e., reference values). All parameters with length units are expressed in our 3D box unit, which has a total side length of 1 and is centered around the origin. For all the runs, a grid resolution of $N = 1024^3$ is used. It is worth pointing out, already at this point, that the results of varying two of these parameters, $h_R$ and $\eta_R$, will not be further discussed in this paper, as our rigorous analysis has clearly shown that they do not affect the power spectrum in any significant way. The parameter study was performed fully on an isotropic field, such as the one shown in Figure 1, where we took all reference values as in Table 1. Limiting the parameter

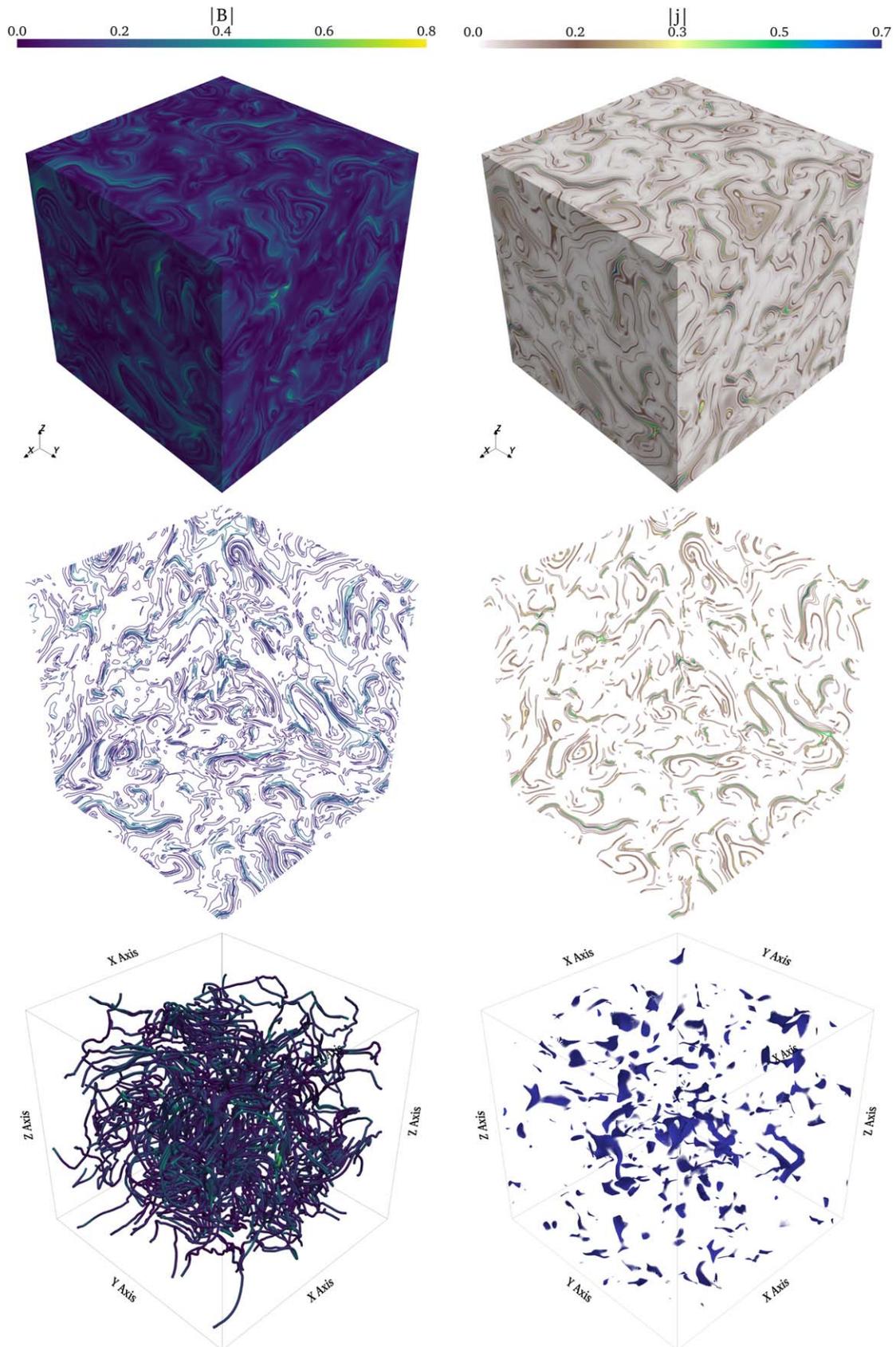

**Figure 1.** Visual representation of a typical turbulent magnetic field (left panels) and current density field (right panels) as generated by BxC. Top row: 3D plot of the norm of the fields. Middle row: 2D cuts of isocontours for different levels ($|\boldsymbol{B}|, |\boldsymbol{j}| = 0.1, 0.2, 0.3, 0.5, 0.7, 0.8$). Bottom row: magnetic field streamlines (left) and current density isocontours (right) for $|\boldsymbol{j}| = 0.6$.

**Table 1**
List of Input Parameters Considered in the Study Together with Their Reference Values and Range of Variation

| Input Parameter | Reference Value | Range of Values |
| --- | --- | --- |
| $L_R$ | 0.15 | [0.05, 0.2] |
| $L_B$ | 0.1 | [0.05, 0.16] |
| $h_r$ | 0.1 | [0.05, 2.0] |
| $h_B$ | 2.0 | [1.0, 3.0] |
| $\eta_r$ | 0.001 | [0.001, 0.01] |
| $\eta_B$ | 0.003 | [0.001, 0.01] |

**Note.** $L_R$, $L_B$, $\eta_r$, and $\eta B$ are all expressed in box units; $h_r$ and $h_B$ are dimensionless quantities.

study to isotropic fields allowed us to isolate the effects of the input parameters from variations caused by anisotropic features.

In order to consistently assess the effect of varying each input parameter, we first identify a set of characteristic features of the power spectrum, highlighted schematically in Figure 2. For a generic turbulent power spectrum, we focus on three points of interest in the $(k, P)$-plane (i.e., wavenumber versus spectral power): the point at which the inertial range begins, the point at which it ends, and the maximum wavenumber (i.e., the point at which the power spectrum ends). The subscripts $i$, $d$, and max stand for injection, dissipation, and maximum, respectively, and are used to distinguish the three different points. Moreover, the spectral slope in the inertial range, which we indicate with $\zeta$, is an important feature that is taken into consideration in this analysis. Among the complete set of features shown in Figure 2, we selected particularly $P_i$, $k_i$, $k_d$, and $\zeta$. The analysis led us to identify a series of relations between the input parameters and the desired spectral feature. For each of these relations, a fit was performed for the purpose of user-friendly customization of the fields. Explicit equations for each fit can be found at the end of this section in Table 2, together with the relative maximum absolute and relative error.

### 3.1. Controlling the Injection Scale and Spectral Power through $L_R$ and $L_B$

The input parameters $L_R$ and $L_B$ control the integration region of the FGF $R$ and the magnetic field $B$, respectively, as described by Equations (2) and (3). It is intuitive to expect a relation between the size of these integration regions and the large-scale structures of the fields. Focusing on the power spectrum features, the large-scale structures are related to the injection scale, i.e., the point at which the power-law behavior of the power spectrum begins. In the context of our analysis, this means that the parameters $L_R$ and $L_B$ are expected to affect the quantities $k_i$ (scale) and $P_i$ (energy) as shown in Figure 2.

Figure 3 shows a parameter scan obtained by varying $L_R$ between 0.05 and 0.2. The top panel shows the power spectra for each value of $L_R$ considered in the analysis. Here, the dashed vertical lines indicate the injection scale $k_i$, and the coloring scheme is done according to the parameter values. As expected, we can control the beginning of the inertial range, causing a shift to the left (i.e., to larger scales) as we increase the parameter $L_R$. A fit performed to quantify the influence of this parameter on the injection scale is shown in the bottom panel of Figure 3. The relationship between $L_R$ and $k_i$ is described accurately by a linear function. It can be noted in this figure that the same $k_i$ might correspond to multiple

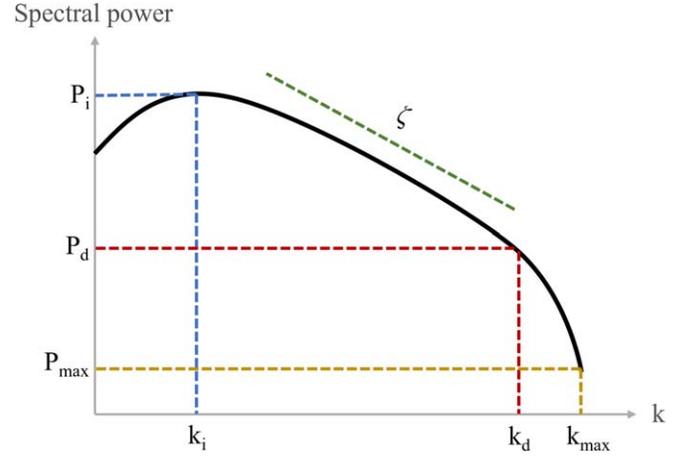

**Figure 2.** Schematic representation of a turbulent power spectrum and its characteristic features. $k_i$, $P_i$ (blue lines) are the wavenumber and spectral power coordinates of the injection scale; $k_d$, $P_d$ (red lines) are the wavenumber and spectral power coordinates of the dissipation scale; and $k_{max}$, $P_{max}$ (yellow lines) are the wavenumber and spectral power coordinates of the maximum point. The slope in the inertial range ($\zeta$) is indicated in green.

**Table 2**
List of All Fitting Curves Obtained for Each Characteristic Feature of the Power Spectrum, with Related Maximum Absolute and Relative Error

| Fit | Relative Error (max) |
| --- | --- |
| $k_i(L_R) \approx -68.5 L_R + 16.4$ | 0.18 |
| $k_d(L_R) \approx \frac{1}{\sqrt{2\pi}\, 0.03} \exp\left(-\frac{(L_R - 0.09)^2}{2(0.03)^2}\right)$ | 0.058 |
| $k_d(\eta_B) \approx 6.2 \times 10^5 \eta_B^2 - 1.2 \times 10^4 \eta_B + 94$ | 0.037 |
| $P_i(L_B; *) \approx -42.6 L_B^2 + 35.3 L_B - 0.9$ | 0.03 |
| $P_i(h_B; *) \approx 1.5 \times 10^4 e^{-4.5 h_B}$ | 0.97 |
| $P_i(\eta_B; *) \approx 6.2 \times 10^4 \eta_B^2 + 18 \eta_B + 0.5$ | 0.1 |
| $\zeta(h_B; *) \approx -0.9 h_B^2 + 6.4 h_B - 10.1$ | 1.94 |
| $\zeta(\eta_B; *) \approx -201.5 \eta_B - 0.5$ | 0.064 |
| $A(\eta_B) \approx -4.6 \times 10^3 \eta_B^2 + 1.13 \times 10^2 \eta_B - 1.2$ | 0.01 |
| $B(\eta_B) \approx 2.1 \times 10^4 \eta_B^2 - 5.2 \times 10^2 \eta_B + 7.7$ | 0.006 |
| $C(\eta_B) \approx -2.2 \times 10^4 \eta_B^2 + 3.7 \times 10^2 \eta_B - 11$ | 0.0038 |

**Note.** The notation (par;*), where par = $L_B$, $h_B$, or $\eta_B$, is used when a feature does not depend on "par" only, but the other parameters on which it depends are kept constant to the reference value.

consecutive values of $L_R$. This aspect is related to the discretization of the domain and the fact that the injection scale is generally found at large scales, where fewer wavenumber points are sampled. Exploring this point is beyond the scope of this work, considering that the maximum relative error on the fit is ≈18% and will be pursued elsewhere. However, we also considered the correlation length $k_c$ as an alternative quantity of interest that could be less affected by the domain discretization. The results of this analysis are shown in Figure 4. As expected, there are fewer overlaps when considering the correlation length. However, the relationship is no longer linear but follows a Gaussian curve. The parameter for the fit, together with the maximum relative error, can be found in Table 2.

Figure 5 shows the results obtained for variations in the $L_B$ parameter. Also here, the top panel shows the power spectra colored according to different values of the parameter, while the same-color dashed horizontal line indicates the corresponding energy at the injection scale. As expected, also in this case we have a clear effect on the injection point: while the scale at which the inertial range begins is not changed, the energy related to the injection scale increases as we consider larger values of $L_B$. The bottom panel of Figure 5 shows a quadratic fit for the relation between $L_B$ and $P_i$.

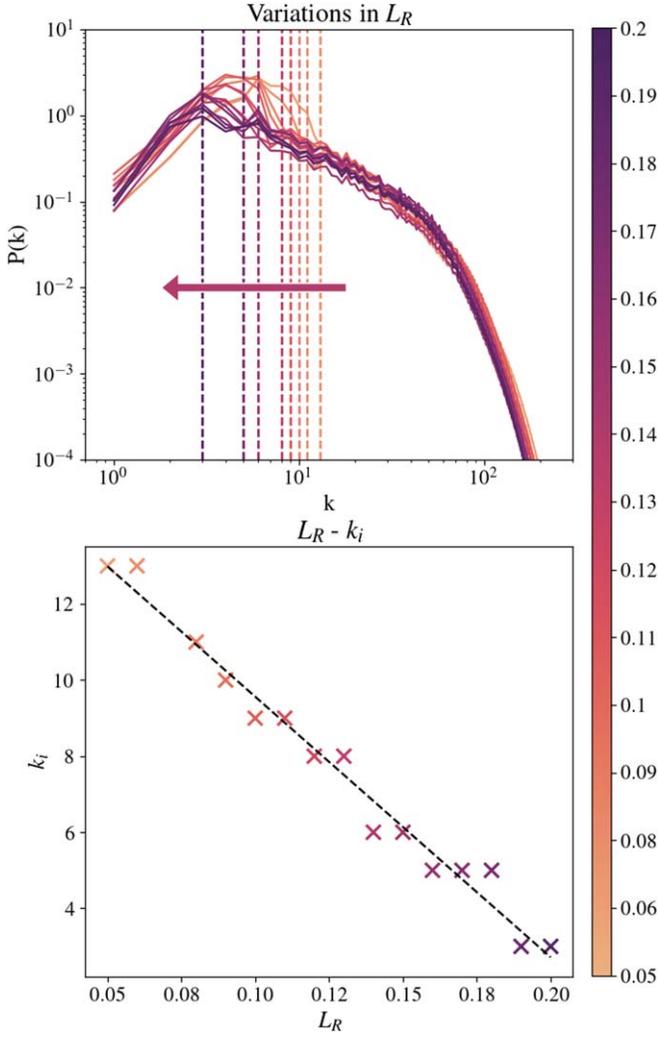

**Figure 3.** Effects of varying the $L_R$ parameter on $k_i$. Top panel: power spectrum for each value of $L_R$; dashed vertical lines correspond to the (variable) injection scale. Bottom panel: fit performed that shows a linear relation between $L_R$ and $k_i$.

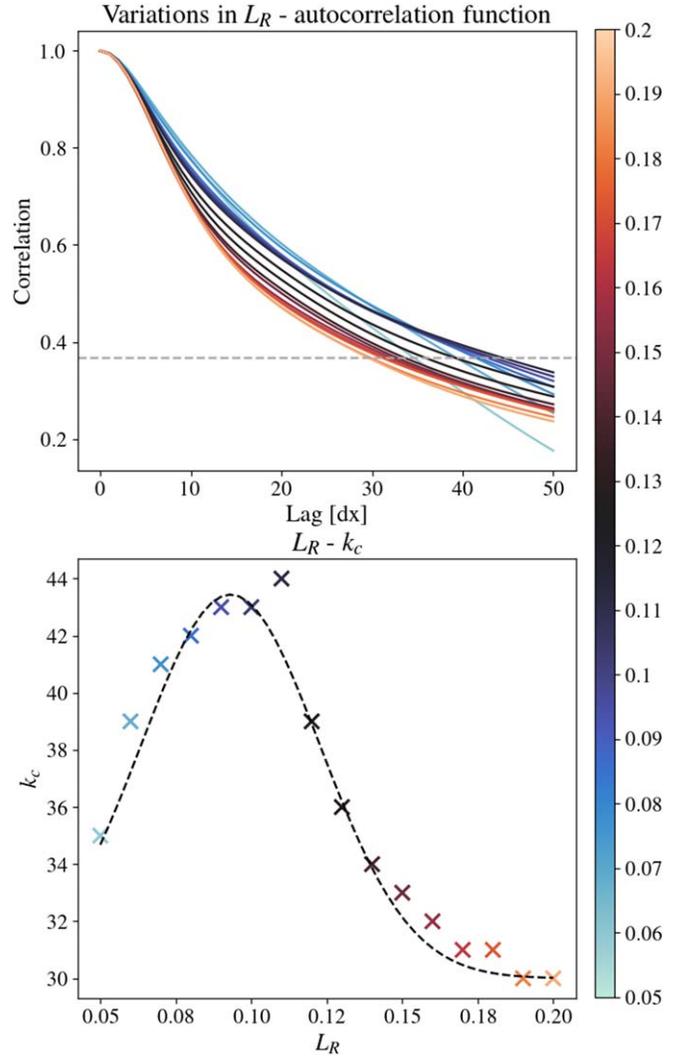

**Figure 4.** Effects of varying the $L_R$ parameter on $k_c$. Top panel: autocorrelation function for each value of $L_R$. Bottom panel: fit performed that shows a Gaussian relation between $L_R$ and $k_c$.

### 3.2. Controlling the Slope of the Power Spectrum through $h_B$

The parameter $h_B$ appears in the exponent of Equation (3), which then presents a spectrum with power-law-like behavior. Therefore, we expect such a parameter to control the spectral exponent of the power law. The results of varying the $h_B$ parameter confirm our expectations and are shown in Figure 6. The top panel shows the power spectra colored according to the parameter values. Here the dashed lines indicate the minimum and maximum slopes, corresponding to the lowest and highest values of $h_B$, respectively. Also in this case, a fit was performed in order to find a specific quantitative relation that describes the dependence of $\zeta$ on the parameter $h_B$. Such a relation is found to be quadratic, as shown in the bottom panel of Figure 6. Moreover, additional runs with higher values of $h_B$ have been performed (reaching $h_B = 4$), which indicates that the slope saturates at roughly $\zeta = 0.7$, making $h_B = 3$ an effective maximum value to be considered.

In the analysis for the $L_R$ and $L_B$ parameters, one can notice that the injection scale and spectral power are the only aspects of the power spectrum affected by the parameter change. On the other hand, when varying $h_B$, changes in the slope are accompanied by changes related to the injection and dissipation scale as well. In particular, changing $h_B$ influences the spectral power at both the injection and the dissipation points, but the relative wavenumbers do not change. For instance, the injection spectral power decreases as the parameter $h_B$ increases, while the injection scale stays constant for all runs. The top panel of Figure 7 shows the fit performed to find the relation between $h_B$ and $P_i$. As indicated in the figure as well, the fit for this relation was performed using only values $h_B \geqslant 1.4$. Initial fits were performed on the entire set of $h_B$ values, considering quadratic and exponential fitting curves. In both cases, the resulting fit was poor, amounting to a maximum relative error of $\approx 77$ and

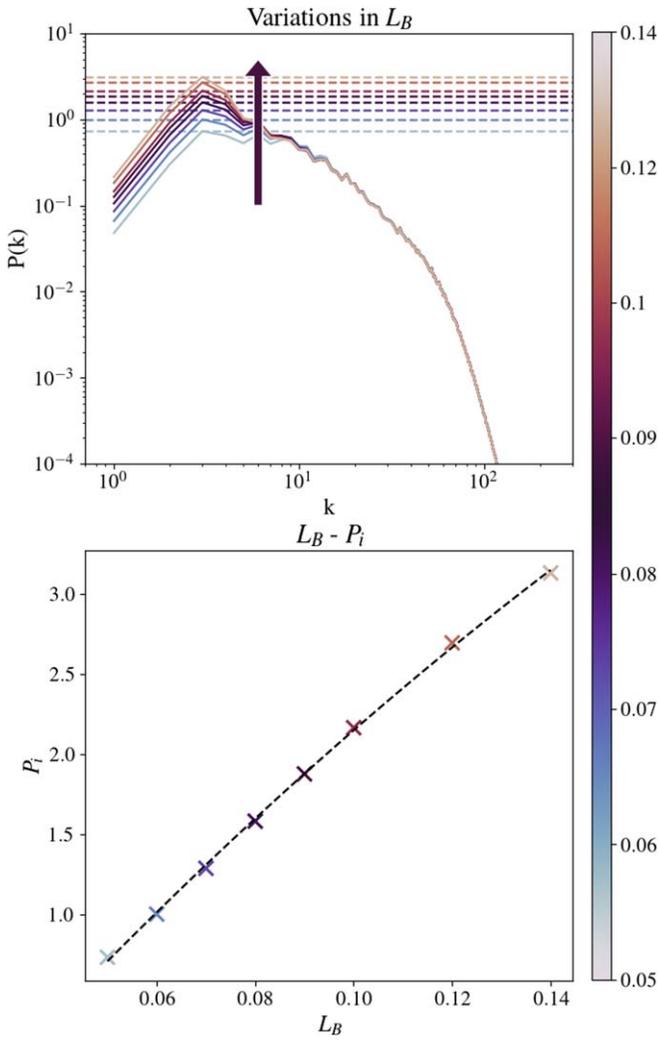

**Figure 5.** Results for varying the $L_B$ parameter. Top panel: power spectrum for each value of $L_B$; dashed horizontal lines correspond to the (variable) injection spectral power. Bottom panel: fits performed to find a quantitative relation between $L_B$ and $P_i$.

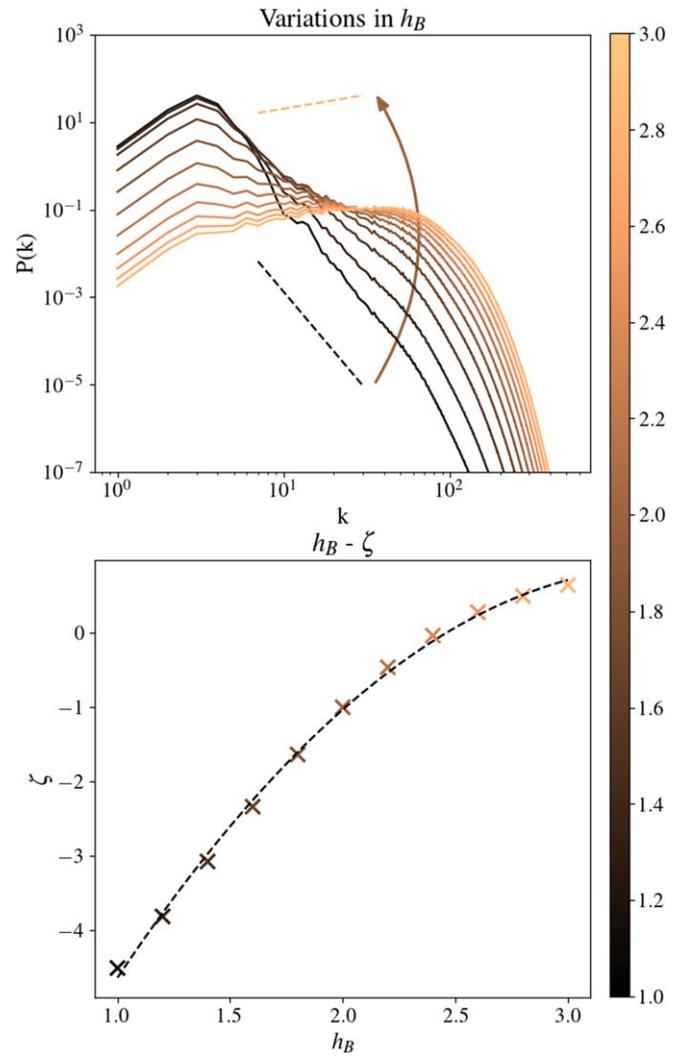

**Figure 6.** Results for varying the $h_B$ parameter. Top panel: power spectrum for each value of $h_B$; dashed lines indicate the slope given by the minimum and maximum spectral index. Bottom panel: fit performed to find a quantitative relation between $h_B$ and $\zeta$. Both plots are colored according to the values of the parameter.

≈23, respectively. In addition to finding a proper fitting relation, we recall that visual aspects of the generated fields have also been taken into consideration. For small values of $h_B$ (i.e., $h_B = 1.0, 1.2$), the fields lose the characteristic shapes and structures that are typical of turbulence. This is clearly visible in the middle panel of Figure 7, in which a 3D box of the norm of **B** is shown for $h_B = 1.0$. The field lacks the typical visual appearance of turbulent fields, and it more closely resembles a Gaussian field, in contrast to the field shown in Figure 1. In this regard, the visual aspect is supported by a statistical analysis as well. The bottom panel of Figure 7 shows the comparison between the PDF of increments in **B** for $h_B = 1$ and $h_B = 2.0$ (i.e., our reference value). The distribution related to $h_B = 2.0$ strongly deviates from a Gaussian distribution, which is in contrast to the behavior exhibited for $h_B = 1$, suggesting that the latter case is not intermittent. In addition, the extremely steep shape of the power spectrum would suggest that the field generated with $h_B = 1$ is not turbulent. All of the above considerations led us to exclude lower values of $h_B$ and consider only those values that generate magnetic fields that can be classified as turbulent from both a visual and a statistical perspective. However, it is worth pointing out that a field with the same characteristics (i.e., turbulence mainly present at large scales) has been recently used by Pezzi & Blasi (2024) in the context of cosmic-ray transport. Similarly, it can be noted that the analysis conducted on the turbulent character of fields generated by low $h_B$ values might also suggest the possibility of having a user-controllable level of intermittency. This aspect is left for future studies.

### 3.3. Controlling the Dissipation Scale and Spectral Power through $\eta_B$

The last parameter considered in this study is $\eta_B$, which is the regularization parameter in the construction of the magnetic field in Equation (3). From a physical and mathematical point of view, $\eta_B$ determines the relative size of the small scales in our system, i.e., the dissipation scales.

The results of varying $\eta_B$ are shown in Figure 8. The top panel shows the power spectra for each value of the parameter considered. Here, the vertical dashed lines represent the dissipation scale, and increasing the value of $\eta_B$ causes a shift toward larger scales. The fit shown in the bottom panel of Figure 8 clearly shows that the relationship between $\eta_B$ and $k_d$ is very well described by a quadratic curve. Looking at the power spectrum overview, one can easily notice that the dissipation scale is not the only aspect of the power spectrum that is affected, since we clearly see that $\eta_B$ affects the spectral slope as well as the injection spectral power, while still leaving the injection scale $k_i$ unchanged. Hence, $\eta_B$ effectively controls both dissipation scale $k_d$ and energy at once. Although the feature $P_d$ is not actively discussed in this work, from the top panel of Figure 8, it is visible that as the dissipation scale $k_d$ moves toward larger scales, the dissipation spectral power $P_d$ decreases.

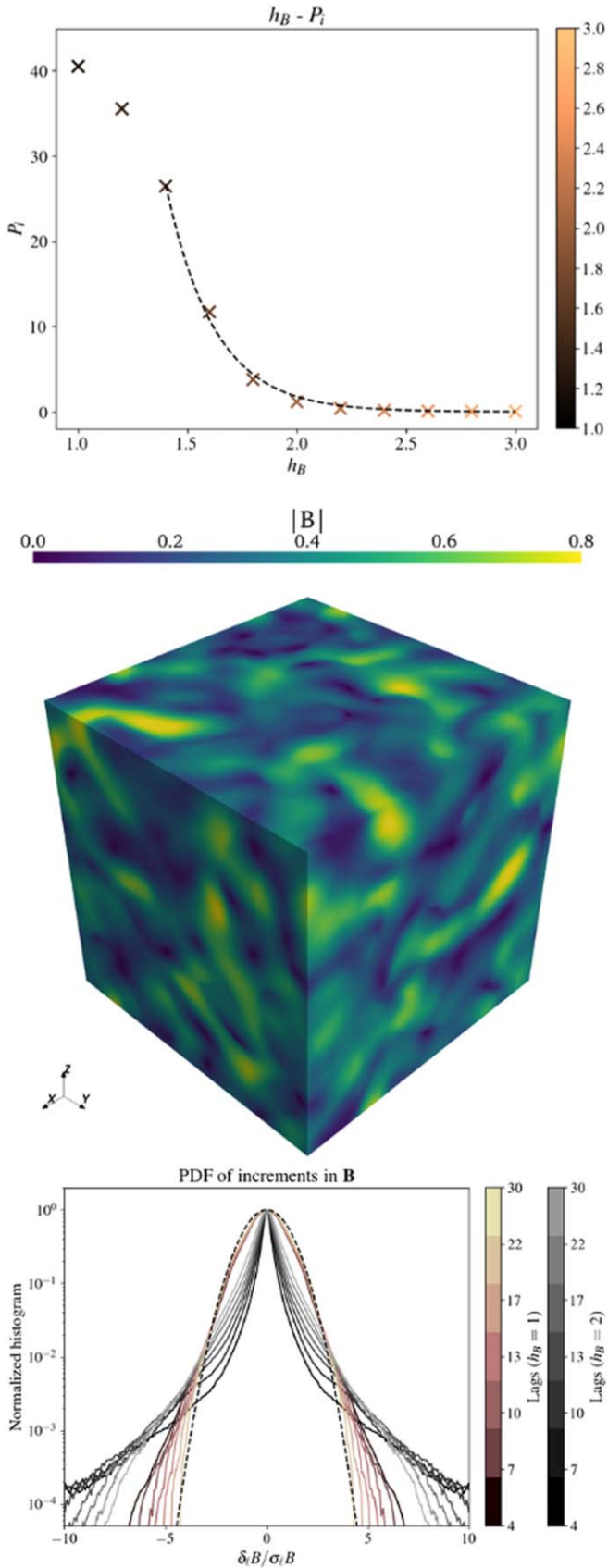

**Figure 7.** Top panel: fit performed to find a quantitative relation between $h_B$ and $P_i$, which is a secondary effect of varying $h_B$. Middle panel: 3D norm of $\boldsymbol{B}$ generated for $h_B = 1$. Bottom panel: comparison between PDF of increments in $\boldsymbol{B}$ for $h_B = 1$ (pink scale) and $h_B = 2$ (gray scale) for different lags.

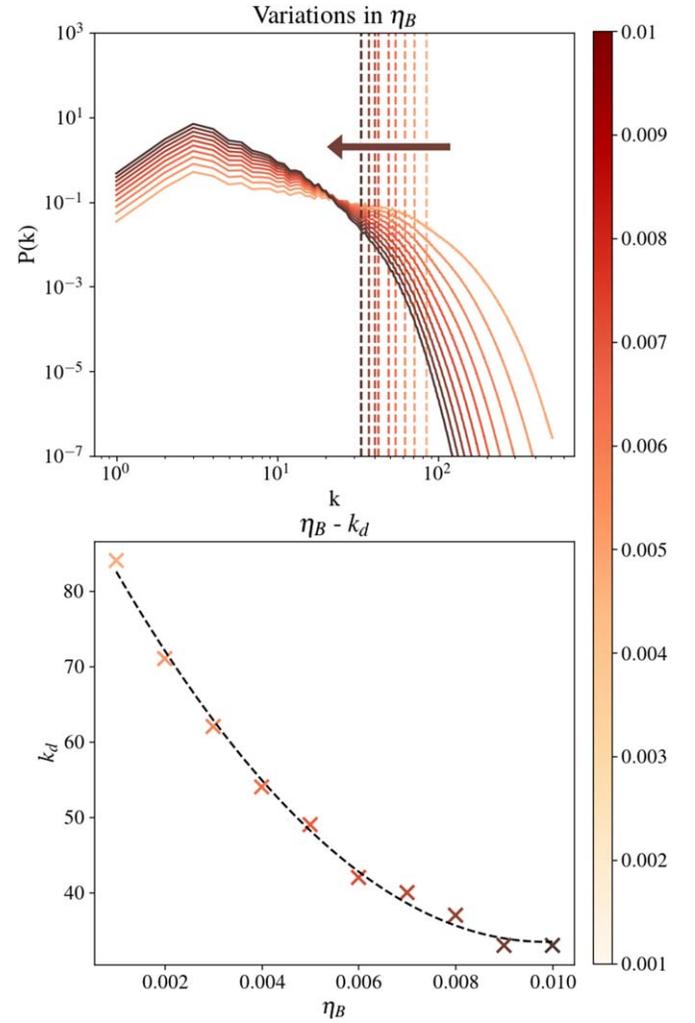

**Figure 8.** Results for varying the $\eta_B$ parameter. Top panel: power spectrum for each value of $\eta_B$; dashed vertical lines correspond to the (variable) dissipation scale. Bottom panel: fit performed to find a quantitative relation between $\eta_B$ and $k_d$. Both plots are colored according to the values of the parameter.

As we are mostly interested in the spectral slope and injection scale aspects, we performed a fit for both the slope $\zeta$ and the injection spectral power level $P_i$ variations. The results are shown in Figure 9. The top panel shows the relation between $\eta_B$ and $\zeta$, which, with a maximum relative error of 0.064, is in good agreement with a linear fit. The bottom panel shows the relation between $\eta_B$ and $P_i$, which in this case is quadratic.

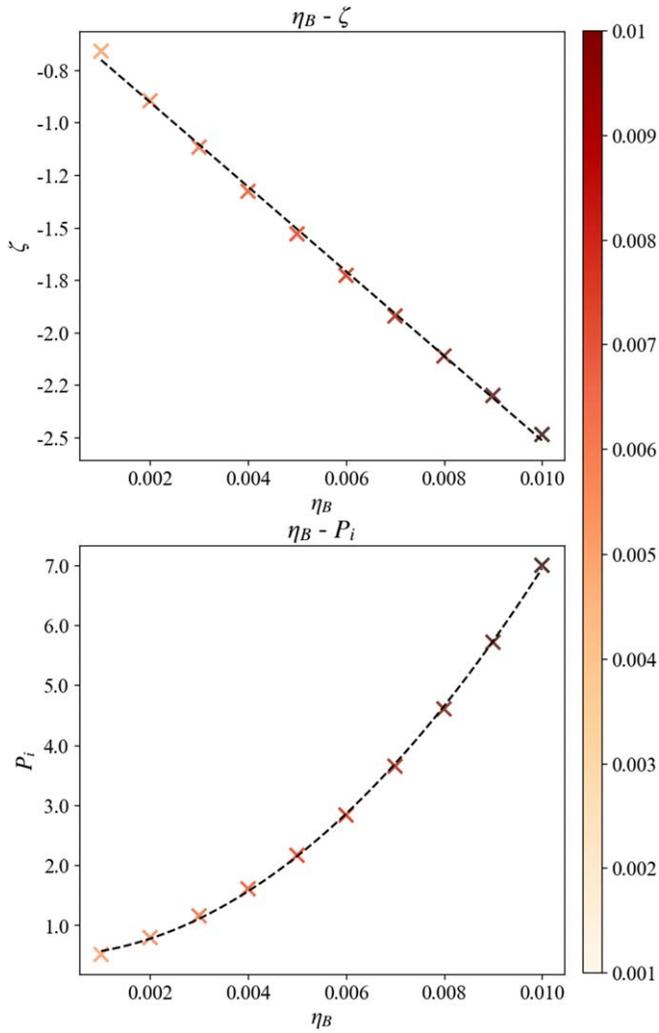

**Figure 9.** Fits performed to find quantitative relations for the secondary effects of changing $\eta_B$. Top panel: relation between $\eta_B$ and $\zeta$. Bottom panel: relation between $\eta_B$ and $P_i$.

### 3.4. Interplay between Parameters

The analysis conducted in the previous sections clearly shows that there is an interplay between parameters, especially with regard to $h_B$ and $\eta_B$. Concerning the effects of changing $\eta_B$, it was previously mentioned that one of the objectives of this study is to be able to control the dissipation scale without affecting either the slope or the injection spectral power of a given power spectrum. The analysis above showed that this is not achievable by only changing the parameter $\eta_B$. However, it is still possible to obtain the same result using a combination of $\eta_B$ and $h_B$, since we found that the effects on the slope and injection spectral power caused by these two parameters are opposite with respect to each other. This means that changing $\eta_B$, while also changing $h_B$ at the same time so that it forces the slope to be constant to the original value, will indeed result in a power spectrum that varies only in the dissipation scale. For this reason, a series of runs was performed varying both $\eta_B$ and $h_B$.

Figure 10 shows the fitting curves of the relationship between $h_B$ and $\zeta$ for different values of $\eta_B$. All the curves are quadratic, but there is a clear pattern dependent on $\eta_B$: as the latter increases, $\zeta$ tends to smaller values. In order to obtain an expression that could relate the slope to both parameters, fits have been performed on the coefficients of the $h_B$–$\zeta$ relations. Eventually, we obtain a relation of the form

$$\zeta(h_B, \eta_B) = A(\eta_B)h_B^2 + B(\eta_B)h_B + C(\eta_B),$$

in which the coefficients $A(\eta_B)$, $B(\eta_B)$, and $C(\eta_B)$ are all quadratic expressions in $\eta_B$.

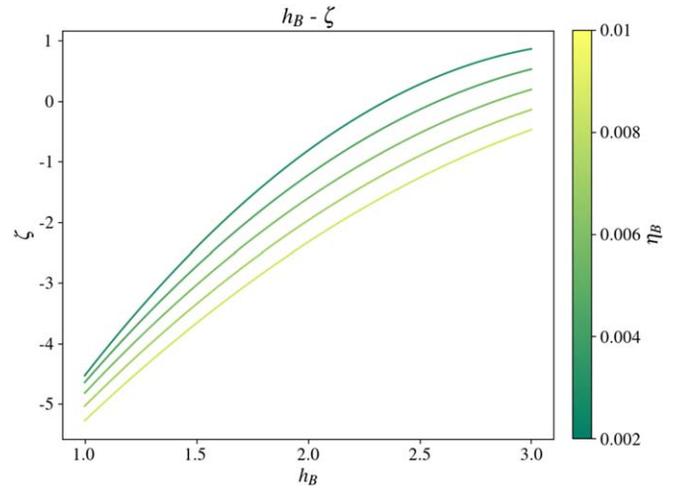

**Figure 10.** Fitting curves describing the relation between $h_B$ and $\zeta$ for different values of $\eta_B$.

## 4. Novel Features

### 4.1. Background Magnetic Fields

One of the main purposes of this work is to extend BxC to reproduce more realistic scenarios, allowing for structures more complicated than purely isotropic turbulent magnetic fields. Indeed, in most actual physical situations, turbulence develops on top of an existing background magnetic field. In this section, a novel method to generate turbulent magnetic fields with arbitrary background structures is presented.

Keeping in mind possible applications of BxC to astrophysics, a set of demonstrative background magnetic field topologies has been implemented in the BxC toolkit. An arbitrary number of additional, user-defined background fields can be easily added to the code. In order of increasing complexity, we now allow for a background uniform magnetic field, a magnetic arcade (with or without shear; Terradas et al. 2015), and a cylindrical flux tube with a force-free Gold–Hoyle model (Vandas & Romashets 2017). The topology of the implemented background field is shown in the left column of Figure 11. The uniform background magnetic field, shown in the upper-left panel of Figure 11, is very simple but represents a variety of physical scenarios that are extremely relevant in the context of astrophysical plasmas, since turbulence is often found and studied in relation to an existing guide field (e.g., Roytershteyn et al. 2015; Dong et al. 2022). The arcade and the flux-rope structure have been chosen as possible models of interest, since the divergence-free nature of magnetic fields implies that flux tubes are cornerstone ingredients in any 3D setting, while arcades occur in many plasma setups from laboratory, over solar, to astrophysical scenarios (e.g., Cheng & Cjoe 2001; Xia et al. 2012; Ryutova 2018).

The general principle on which this method is based is to first generate the background magnetic field $\boldsymbol{B}_0$ independently of the turbulent, BxC-generated field $\boldsymbol{B}_{\mathrm{turb}}$, with the only restriction that both of them have to be defined on the same grid and that the background field must (analytically) be divergence-free. Eventually these two fields, $\boldsymbol{B}_0$ and $\boldsymbol{B}_{\mathrm{turb}}$, are summed up pointwise, obtaining a turbulent magnetic field $\boldsymbol{B}_{\mathrm{str}}$ with a background structure. In the implemented model, both fields are normalized to the maximum value of their norm, leading to fields with a maximum value of order unity. At the same time, the purely turbulent magnetic field $\boldsymbol{B}_{\mathrm{turb}}$ is built using a filtering function in order to ensure that the background structure does not lose its characteristic geometry due to excessively strong turbulent fluctuations. In our examples, such a filtering function is chosen to be the norm of the background magnetic field $\Gamma = |\boldsymbol{B}_0|$; hence, the intensity of the turbulent field is modulated according to the pointwise value of $|\boldsymbol{B}_0|$. This yields turbulent fields with lower intensity at the points in which the background magnetic field is weaker, thus allowing for the field topology to still be relevant. The filtering function $\Gamma$ is implemented inside the modified Biot–Savart law for the magnetic field, meaning that Equation (3) becomes

$$\boldsymbol{B}_{\mathrm{turb}} \propto \int_{r \leqslant L_B} \frac{\Gamma \boldsymbol{c} \times \boldsymbol{r}}{(r^2 + \eta_B^2)^{h_B}} dV. \qquad (7)$$

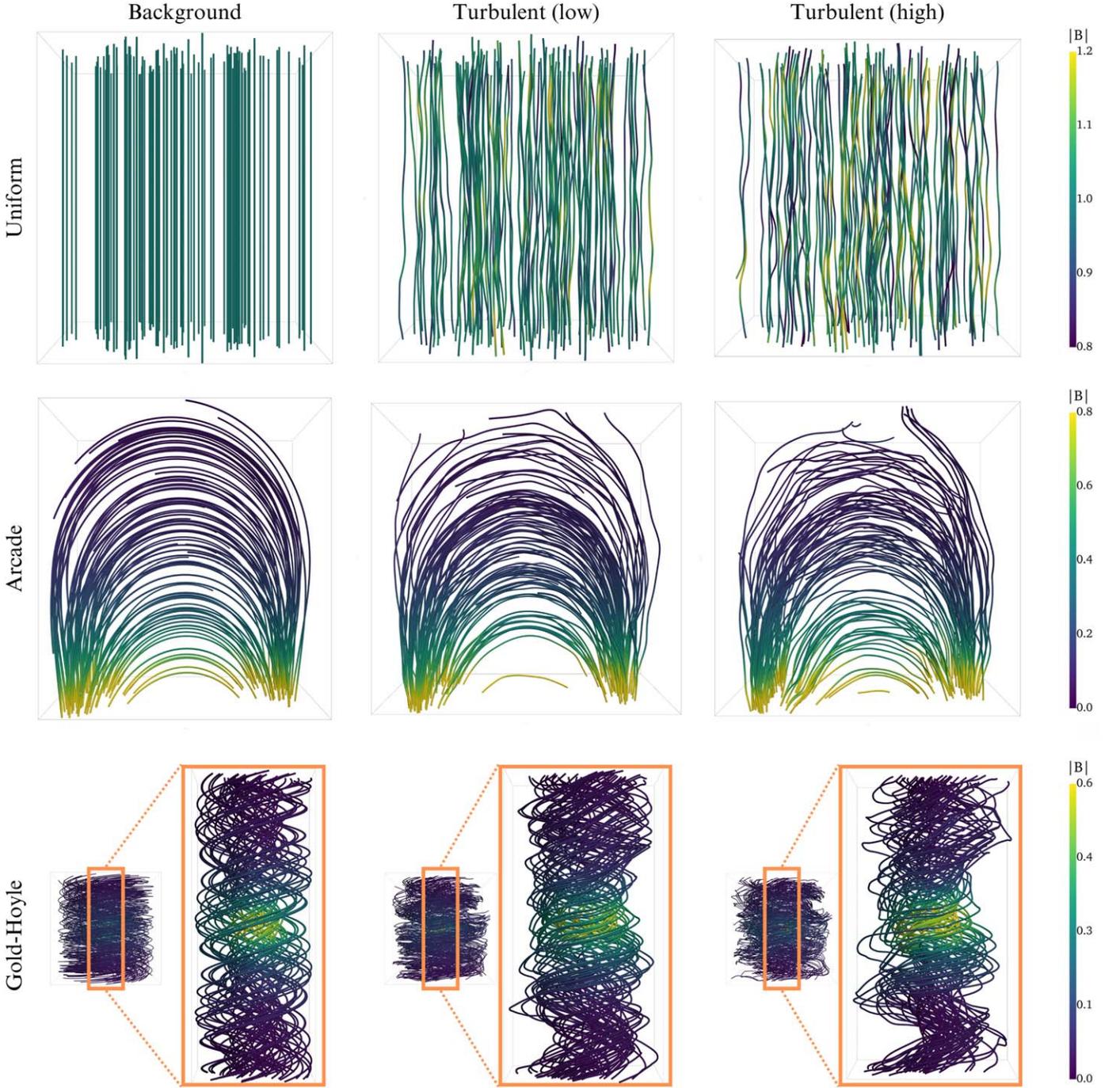

**Figure 11.** Background magnetic field structures implemented in the model and their turbulent realizations for different intensities of turbulence. Top row: uniform magnetic field; middle row: arcade with shear (Terradas et al. 2015); bottom row: cylindrical flux tube with a force-free Gold–Hoyle model (Vandas & Romashets 2017). Left column: background structures; middle column: lightly turbulent realizations; right column: highly turbulent realizations.

Implementing the filtering function inside the modified Biot–Savart law is an important step, necessary to preserve the divergence-free property of the fields. It can be proven (see the Appendix) that the divergence-free property of the (modified) Biot–Savart law is independent of $c$, due to properties of the convolution. In practice, this allows us to multiply $c$ by any scalar field and still generate fields that are analytically divergence-free. Eventually, the final formula to obtain a turbulent magnetic field with the structure of a background magnetic field is

$$\boldsymbol{B}_{\rm str} = \boldsymbol{B}_0 + \gamma \boldsymbol{B}_{\rm turb}(\Gamma), \qquad (8)$$

where $\gamma$ is simply a constant value that gives the user the possibility to regulate the intensity of turbulence with respect to the background magnetic field in addition to (and independently from) the scaling/normalization performed inside the integral. In contrast to the pointwise scaling performed by $\Gamma$, which deals with the relative intensity of the turbulent field within the box itself, $\gamma$ scales the entire turbulent field so as to regulate its intensity compared to the background field. Figure 11 shows turbulent magnetic fields with each background structure implemented in the model for different values of $\gamma$. The middle column shows lightly turbulent fields, while the right column shows highly turbulent fields, obtained for low and high values of $\gamma$, respectively.

### 4.2. Anisotropy

As described in Section 2, our BxC algorithm can use direction-dependent model parameters for generating anisotropic fields. The introduction of anisotropy in the fields is an important feature in order to have realistic synthetic turbulent fields that users can customize according to the application. In numerous astrophysical plasmas, turbulence is found to develop anisotropically, meaning that there exists a preferential direction with respect to a background (or guide) field modifying energy transfer along parallel and perpendicular directions from large to small scales. Such anisotropy is still heavily researched in magnetic field turbulence (e.g., Montgomery & Turner 1981; Matthaeus et al. 1996; Müller et al. 2003; Horbury et al. 2008), and precise scaling laws have not been determined in general. The most well-studied case is known as "critical balance" (CB; Sridhar & Goldreich 1994; Goldreich & Sridhar 1995, 1997), for which scaling laws for parallel and perpendicular components of the field have been experimentally determined and validated with physical models (see, e.g., Oughton & Matthaeus 2020 for an overview of CB theory and its applications). For this reason, our BxC tool has been extended to provide the freedom of obtaining different scalings in the parallel and perpendicular directions. However, the feature of anisotropy will be discussed here relatively qualitatively, as more dedicated work is currently required to quantify the precise effect of the numerical value of all parameters.

The approach used to obtain anisotropic fields is the following. First, we identified the parameters in the algorithm that could potentially induce differences in the resulting fields based on direction. Second, these parameters were generalized from a single value for all directions to a direction-dependent three-value implementation. Lastly, different runs were made and visualized to test that the generated fields were effectively anisotropic. The analysis of the generated field was conducted from both a visual point of view and a statistical one. Visually, the aim was to reproduce turbulent structures, particularly current sheets, that are aligned with a preferential direction. This is often found in simulations that reproduce turbulent magnetic fields in the presence of a guide field (e.g., Roytershteyn et al. 2015; Dong et al. 2022). For the statistical analysis, we compute the power spectrum in its parallel and perpendicular component, where the parallel direction is assumed to be along $\hat{z}$. After computing the 3D power spectrum $P(k_x, k_y, k_z)$, we integrate in the $(k_x$–$k_y)$ plane summing over concentric shells as follows:

$$P(k_\perp, k_z) = \sum_{0 \leqslant \sqrt{k_x^2 + k_y^2} < k_{\rm max}} P(k_x, k_y, k_z),$$

where $k_{\rm max}$ is defined as the miminum between $\max(k_x)$ and $\max(k_y)$. One can then directly visualize the 2D spectrum. The result of this study are shown in Figure 12 for the different cases considered, together with the isotropic case (first column) for easier comparison.

The first parameter to be considered for controlling and introducing anisotropy is the standard deviation of the white-noise vector. For the isotropic case, this vector field was initialized from a normal distribution of zero mean ($\mu = 0$) and unit standard deviation ($\sigma = 1$) in all three directions. Allowing $\sigma$ to have different values for $x$, $y$, and $z$ has the desired effect of inducing anisotropy. In particular, wanting to mimic the presence of a background field in the $z$-direction, $\sigma_z$ has been increased compared to $\sigma_x$ and $\sigma_y$, which are both set to unity. For what concerns the resulting white-noise vector field, these different values of the standard deviation generate a field that has more energy in its $z$-component. However, this intrinsic difference between the three directions propagates throughout the algorithm and eventually translates into anisotropy between the parallel ($z$-axis) and perpendicular ($x$–$y$ plane) directions. The second row of Figure 12 shows the results of the analysis conducted on a field generated using $\sigma = [1, 1, 10]$. The left panel shows the 1D reduced power spectrum, while the middle panel shows the 2D power spectrum, both featuring parallel and perpendicular components. The black lines in the 2D power spectrum correspond to isocontours of spectral power levels. The asymmetry of the contours (i.e., slight elongation in the parallel direction) is indeed an indication that the spectral power and the associated power-law decay differ between the two directions. The same spectral power level corresponds to larger wavenumbers in the parallel direction than in the perpendicular one. This difference can also be noticed in the 1D power spectra. The right panel shows the current density isocontours for $|\boldsymbol{j}| = 0.35$. This figure gives visual indications that the resulting field features a preferential direction. In fact, the current density sheets are predominantly extended along the $z$-axis, contrary to what was shown for the isotropic case. In the preliminary analysis that was conducted for this case, different values of $\sigma_z$ have been analyzed and compared to each other. Here we show the case that produces the maximum level of anisotropy that was observed.

The second parameter considered to construct anisotropic fields is the integration region of the Biot–Savart law from Equation (3). Although the integration regions for the construction of both the FGF and the magnetic field have been considered in the analysis, only using a vector-valued parameter $L_R$ led to more significant results in terms of anisotropic features. Also for this case, as a preliminary survey, we considered different combinations of $L_{R,x}$, $L_{R,y}$, and $L_{R,z}$. Here we show the two most anisotropic fields observed. The results are shown in the third and fourth rows of Figure 12 for $L_R = [0.4, 0.4, 0.01]$ and $L_R = [0.01, 0.01, 0.4]$, respectively. Recall that we compute the fields on a box of unit length, which makes $L_{R,i} = 0.4$ an effective maximum value to use in this case. Increasing $L_{R,x}$ and $L_{R,y}$ produce results that are similar to the $\sigma_z$ case: the 1D reduced spectra and the 2D spectrum isocontours elongated in the $k_\perp$-direction indicate different spectral power distribution and transfer, and $\boldsymbol{j}$ isocontours clearly elongated in the $z$-direction show the existence of a preferential direction. The results of increasing the $L_{R,z}$ parameters are also similar but opposite: 1D power spectrum shows different powers and spectral slopes, the 2D spectrum isocontours are elongated in the $k_\parallel$-direction, and the $\boldsymbol{j}$ isocontours are elongated in the $x$–$y$ plane. Comparing these results to the ones obtained for $\sigma_z$, one can notice an increased level of anisotropy in the fields. The 2D power spectrum is more skewed, and the current density isocontours follow the preferential direction more closely, assuming an almost bidimensional character. However, we also conducted a preliminary analysis concerning the variance anisotropy. Such quantity is computed as

$$\Sigma = \frac{\langle b_\perp \rangle}{\langle b_\parallel \rangle},$$

where $b_\parallel$ and $b_\perp$ indicate the fluctuations from the mean value in the parallel and perpendicular directions, respectively, and $\langle \rangle$ indicates a spatial average. From this analysis, we concluded that the most efficient way to introduce variance anisotropy is by varying $\sigma_z$. Indeed, $\Sigma$ computed for the cases of $L_R = [0.4, 0.4, 0.01]$ and $L_R = [0.01, 0.01, 0.4]$ shows slight variation with respect to the isotropic case, while the variance anisotropy measured for the $\sigma_z = 10$ case leads to a difference of 81%.

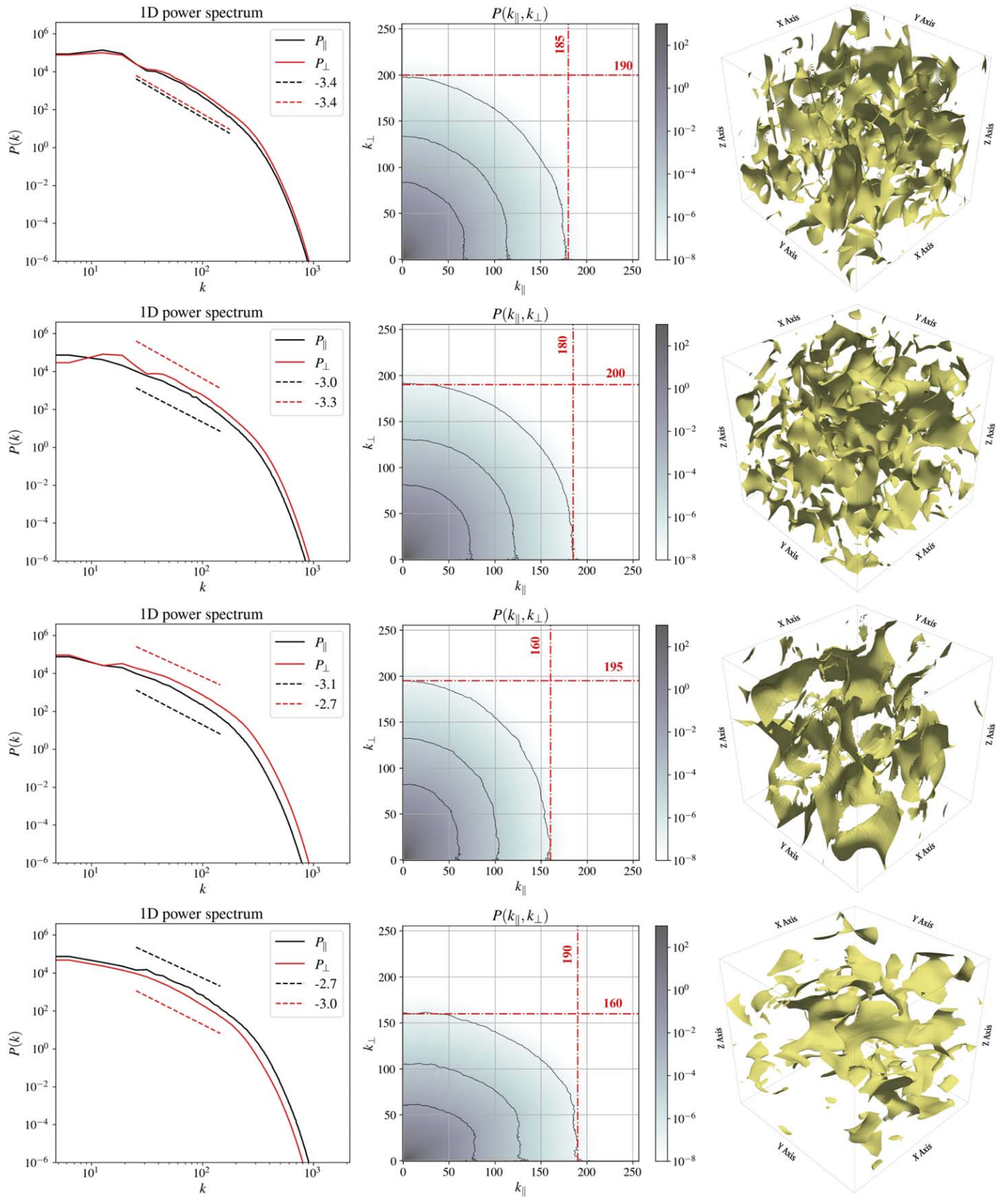

**Figure 12.** Results of anisotropy study in terms of 1D power spectra (left column), 2D power spectra (middle column), and current density isocontours for $|j| = 0.35$ (right column). First row: isotropic case; second row: $\sigma_z = 10$; third row: $L_R = [0.4, 0.4, 0.01]$; fourth row: $L_R = [0.01, 0.01, 0.4]$.

As mentioned at the beginning of this section, this analysis was performed on a rather qualitative level. More in-depth studies of the input parameters should be performed in order to quantify their effect on the resulting anisotropy (both spectral and variance) in the fields. At the same time, this preliminary study effectively shows two techniques through which the BxC-generated fields can be rendered anisotropic.

## 5. Conclusions and Outlook

Synthetic turbulence models represent a promising alternative to and/or support for numerical simulations, having the advantage of requiring few computational resources and time. In this work, we have significantly developed our BxC toolkit, improving control over the existing generated turbulent magnetic fields and implementing novel features that are able to render more realistic fields.

The first part of this work focuses on providing the user with precise information on how to control the power spectrum of the generated fields. To do so, an in-depth parameter study has been performed on the parameters that define the modified Biot–Savart laws. After the identification of possible power spectrum features of interest, it has been shown how variations in a restricted set of input parameters can affect and control such features. In particular, the integration region parameters $L_R$ and $L_B$ control the injection scale in terms of wavenumber and spectral energy, respectively. The free power-law exponent $h_B$ controls the slope of the power spectrum, allowing for a wide range of spectral exponents. The regularization parameter $\eta_B$ controls the dissipation scale. Contrary to what happens at the injection scale, in which the two parameters separately control the two coordinates of the injection point, the regularization parameter affects the dissipation scale in terms of both wavenumber and energy level. This produces a change in the slope of the spectrum and hence an entanglement with the parameter $h_B$. The connection between the two parameters, however, is an advantage that one can exploit in order to produce a change in the dissipation scale only with respect to the wavenumber. This can easily be done by appropriately combining changes in both parameters. The parameter study has been conducted in a systematic way, and quantitative relations between parameters and power spectrum features are provided for each parameter considered in the analysis. Table 2 contains all the relations found through the analysis and represents an important tool to generate isotropic magnetic fields with given spectral properties.

The second part of this study aimed at further improving the generated fields in terms of realistic features that can be implemented. In particular, the development focused on two main aspects: generating turbulent fields that are not only purely turbulent but have a background structure and introducing anisotropy in the fields. For the first point, it has been shown that it is possible to obtain turbulent fields with a background structure by summing an appropriately constructed purely turbulent field to the desired background magnetic field. Here, "appropriately" means that the turbulent field should be modulated in intensity based on the background field, to avoid the disruption of structures where the background field is weak. It is also suggested that the summing process should be modulated by a constant value, which gives users an additional degree of freedom to regulate the intensity of turbulence with respect to the background field.

Finally, the synthetic model was generalized in order to introduce anisotropy in the fields. With the purpose of keeping the model as simple and efficient as possible, anisotropy has been introduced by changing a set of chosen parameters from single-valued to direction-dependent values. Two methods to introduce anisotropy in the fields have been investigated: increasing the values of the white-noise standard deviation in one direction and using different values of the integration region in three directions. Both methods successfully reproduced skewed 2D power spectra and current density isocontours extending along a preferential direction. Among these two, different sizes of the integration region produced more significant and effective results. The analysis presented in this work was conducted on a qualitative level, and the quantification of the precise effects of these parameters is left for further studies.

The quantification of the effects of the input parameters and the novel features introduced in this work represent an important milestone in the development and application of the BxC code. As of now, we can inexpensively generate large data cubes of the turbulent magnetic field and turbulent magnetic field structure, with user-controlled power spectrum features. This paves the way for possible astrophysical applications, such as the study of cosmic-ray propagation (e.g., Pucci et al. 2016; Dundovic et al. 2020; Reichherzer et al. 2020; Kuhlen et al. 2022; or see Mertsch 2020 for a review of test particle simulations of cosmic rays in synthetic turbulent fields). Future development options include user-controlled higher-order statistics (i.e., intermittency), as well as generalizing the model to both flow and magnetic field realizations with combined turbulent properties, which would make BxC suitable for many more applications (e.g., studying particle acceleration, providing initial conditions for DNSs).

## Acknowledgments

D.M. sincerely thanks J. B. Durrive for the contribution and insights on the preexisting model and F. Pucci for the constructive feedback and valuable suggestions. R.K. is supported by Internal Funds KU Leuven through the project C14/19/089 TRACE-Space and an FWO project G0B4521N, along with funding from the European Research Council (ERC) under the European Union Horizon 2020 research and innovation program (grant agreement No. 833251 PROMINENT ERC-ADG 2018). F.B. acknowledges support from the FED-tWIN program (profile Prf-2020-004, project "ENERGY") issued by BELSPO and from the FWO Junior Research Project G020224N granted by the Research Foundation—Flanders (FWO).

We acknowledge support by the VSC (Flemish Supercomputer Center), funded by the Research Foundation—Flanders (FWO) and the Flemish Government—department EWI.

## Appendix
## Proof of Divergence-free Fields

BxC generates synthetic turbulent magnetic fields using a modified version of the Biot–Savart law, as stated in Equation (3). In order to prove the divergence-free property of the fields generated by BxC, let us write the convolution in Equation (3) in components:

$$B_i = N \epsilon_{iab} c_a * k_b. \quad \text{(A1)}$$

Here $\epsilon_{iab}$ is the Levi–Civita symbol, we indicate with $N$ the proportionality constant, and we have implicitly defined

$$k_b = \frac{x_b}{(\|\boldsymbol{x}\|^2 + \eta^2)^h} \Phi_L(\|\boldsymbol{x}\|^2),$$

where $\Phi_L$ is the integration function. Using the property of convolutions for which $(f*g)' = f*g'$, for each component $B_i$, it holds:

$$\partial_i B_i = N \epsilon_{iab} c_a * (\partial_i k_b).$$

Let us now work further on the derivative of $k_b$. In particular, let us compute the derivative of each term ($x_b$, $(x_a x_a + \eta^2)^{-h}$ and $\Phi(x_a x_a)$) separately:

$$\partial_i x_b = \delta_{ib},$$
$$\partial_i (x_a x_a + \eta^2)^{-h} = -h(x_a x_a + \eta^2)^{-h-1} \partial_i(x_a x_a)$$
$$= -h(x_a x_a + \eta^2)^{-h-1} 2 x_a \delta_{ia},$$
$$\partial_i \Phi_L(x_a x_a) = \Phi_L'(x_a x_a) \partial_i(x_a x_a)$$
$$= \Phi_L'(x_a x_a) 2 x_a \delta_{ia}.$$

Combining these results together, we obtain

$$\partial_i k_b = \frac{1}{(\|\boldsymbol{x}\|^2 + \eta^2)^h}$$
$$\times \left( \delta_{ib} \Phi_L(\|\boldsymbol{x}\|^2) - \frac{2h x_i x_b}{\|\boldsymbol{x}\|^2 + \eta^2} + 2 x_i x_b \Phi_L'(\|\boldsymbol{x}\|^2) \right) =: T_{ib}.$$

From the latter equation, one can infer that $T_{ib}$ is symmetric in $i \leftrightarrow b$, as long as $\partial_i \Phi_L \propto x_i$. Once the symmetry of $T_{ib}$ has been proven, one can easily show

$$\delta_i B_i = N \underbrace{\epsilon_{iab}}_{\text{antisymmetric}} c_a * \underbrace{T_{ib}}_{\text{symmetric}} = 0.$$

Since this result holds for all $i$, we have proven that BxC generates divergence-free fields, with constraints only on the integration function.

## ORCID iDs

Daniela Maci  https://orcid.org/0000-0002-3392-8329
Rony Keppens  https://orcid.org/0000-0003-3544-2733
Fabio Bacchini  https://orcid.org/0000-0002-7526-8154